\documentclass{mem}
\usepackage{natbib}
\usepackage{txfonts}
\usepackage{balance}
\usepackage{graphicx}
\usepackage{xcolor}
\usepackage{flushend}
\usepackage[a4paper,breaklinks,pdftex]{hyperref}
\idline{75}{282}
\begin{document}
\def\teff{$T\rm_{eff }$}
\def\kms{$\mathrm {km s}^{-1}$}

\title{
A versatile classification tool for galactic activity using optical and infrared colors
}

   \subtitle{}

\author{
E. \,Kyritsis\inst{1} 
\and C. \, Daoutis\inst{1}
\and A. \, Zezas\inst{1}
\and K. \, Kouroumpatzakis\inst{2,1}
          }

\institute{
          Institute of Astrophysics, FORTH / Dept. of Physics, Univ. of Crete, Heraklion, Greece
         \and
         Astronomical Institute, Ac. of Sciences, Boční II 1401, CZ14131 Prague, Czech Republic\\
         \email{ekyritsis@physics.uoc.gr}
}

\authorrunning{Kyritsis}

\titlerunning{Galactic activity diagnostics}

\date{Received: Day Month Year; Accepted: Day Month Year}

\abstract{
We use the Random Forest (RF) algorithm to develop a tool for automated activity classification of galaxies into 5 different classes: Star-forming (SF), AGN, LINER, Composite, and Passive. We train the algorithm on a combination of mid-IR (WISE) and optical photometric data while the true labels (activity classes) are based on emission line ratios. Our classifier is built to be redshift-agnostic and it is applicable to objects up to z $\sim$0.1. It reaches a completeness $>$80\% for SF and Passive galaxies, and $\sim$60\% for AGN. Applying it to an all-sky galaxy catalog (HECATE) reveals a large population of low-luminosity AGNs outside the AGN locus in the standard mid-IR diagnostics.

}
\maketitle{}

\section{Introduction}
Activity classification of galaxies is of great importance for many fields of extragalactic Astrophysics, such as understanding galaxy evolution \citep{Kewley2019} and/or AGN demographics. Traditionally, this is done using characteristic emission--line ratios which discriminate galaxies into different classes depending on the source of ionization \citep[e.g.][]{Kewley2019}. However, the need for spectroscopic data hampers the applicability of these diagnostics to very large datasets since spectroscopic observations are available for a subset of the objects with photometric data. In addition, these diagnostics cannot be used on galaxies without emission lines rendering them inapplicable to passive galaxies. While alternative  diagnostics based on mid-IR colors \citep[][hereafter M12 and A13]{Mateos2012,Assef2013} are successfully used for identifying luminous AGNs, they are not as reliable in the local universe.

To address these limitations, we develop a new activity diagnostic by combining the RF machine learning algorithm \citep{Louppe2014} with multi-wavelength photometric data.

\section{Classification scheme and data}

\subsection{Photometric data}
\begin{figure}[]
\resizebox{\hsize}{!}{\includegraphics[clip=true]{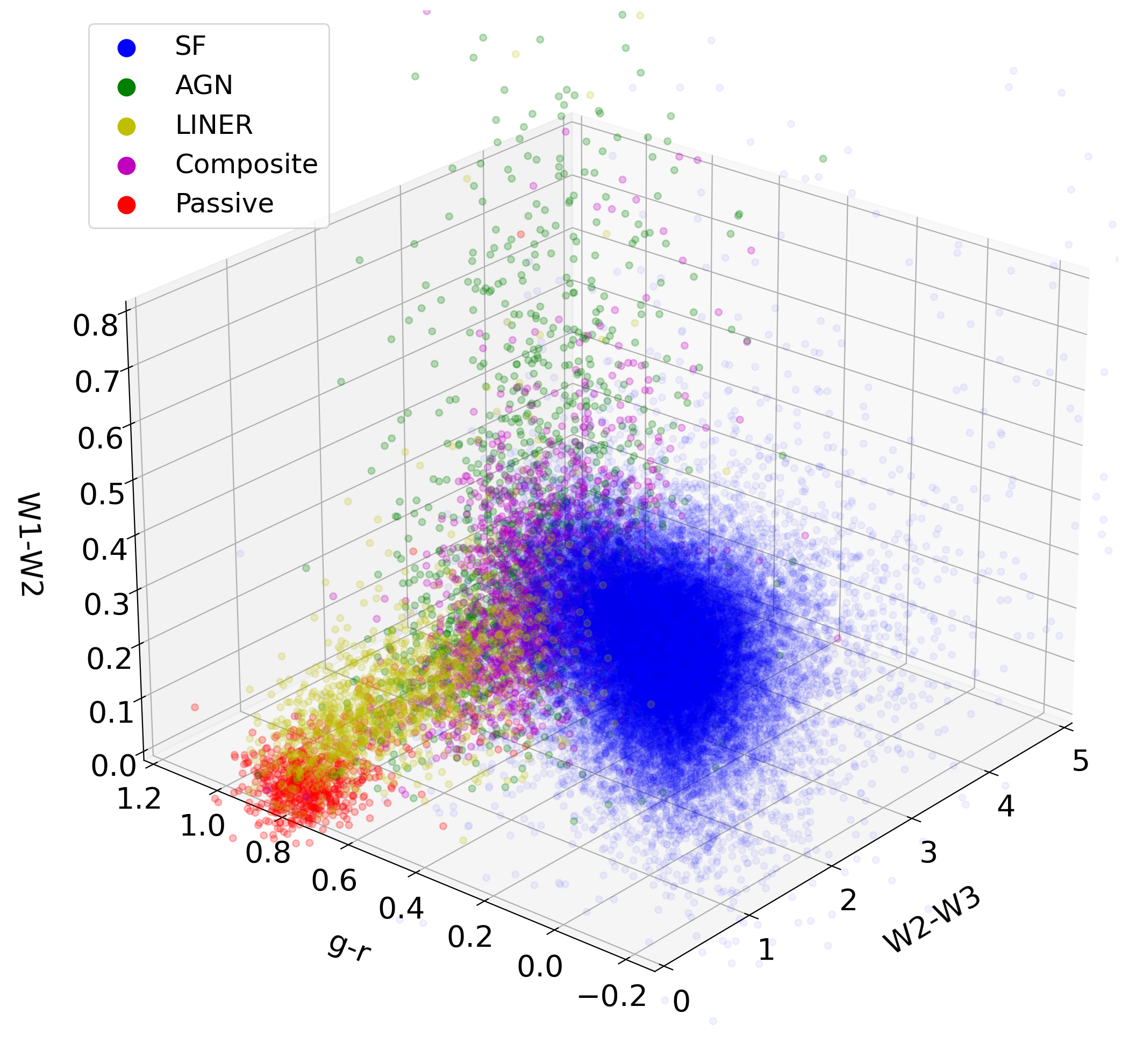}}
\caption{
\footnotesize
Distribution of our training sample in the feature space. The 5 classes of our classification scheme are well separated with a higher mixing between the Composite and AGN activity classes.  
}
\label{fig:1}
\end{figure}

Galaxies have different spectral shapes depending on their source of ionization. Previous works have shown that these differences are stronger in the UV, optical, and mid-IR bands. In order to maximize the available sample, we opted to use for training our algorithm mid-IR and optical photometric data from the AllWISE Source Catalog \citep{Wright2010} and the SDSS DR16 \citep{Brinchmann2004}, respectively. To avoid the need for redshift measurements, we use colors rather than luminosities. In order to mitigate aperture effects, we use a mid-IR hybrid-photometric scheme that combines custom-aperture photometry for nearby (extended) and fixed aperture photometry for more distant point-like sources. The optical data consist of $g-r$ colors, based on SDSS DR16 fiber-photometry. In order to reduce the noise in our training data set we chose only galaxies with S/N $>$ 5 (Signal-to-Noise) for the optical bands g,r and the mid-IR bands W1, W2. For the band W3 we used a S/N $>$ 3. Our final optimal feature scheme comprises colors: W1-W2, W2-W3, and $g-r$ and its distribution per activity class is presented in Fig. \ref{fig:1}.

\subsection{Classifcation scheme} 
We adopt a 5-class classification scheme that discriminates galaxies into different activity classes: SF, AGN, LINER, Composite, and Passive. In order to construct the training sample, we use spectroscopic information from the SDSS-MPA-JHU \citep{Brinchmann2004} catalog by selecting only the galaxies that show strong emission lines (S/N $>$ 5). The emission-line galaxies are classified based on the 4-dimensional data-driven classification algorithm of \cite{Stampoulis2019}. To define a sample of Passive galaxies without emission lines we selected objects with good spectra (continuum S/N$_{cont}$ $>$ 3) and absent emission lines (S/N$_{line}$ $<$ 3). Our final sample includes 40954 galaxies spanning a redshift range of $z$ = 0.02 - 0.08. Table \ref{tab:1} shows the composition of our final sample classification scheme. For the
training of the RF algorithm we considered 50\% of the full set (20477/40954), and for the test the rest 50\%. Given the strong imbalance between the different classes in the training sample, we used a stratified split in order to ensure that both training and the test set have the same proportions of each class.

\begin{table}[]
\caption{The composition of the training sample per galaxy activity class that was used in the training sample.} 
\centering 
\resizebox{\columnwidth}{!}{%
\begin{tabular}{l c c}
\hline\hline
  Class & Number of objects & Percentage (\%)\\
\hline
Star-forming & 35878 & 87.6\\
Seyfert & 1337 & 3.3\\
LINER & 1322 &  3.2\\
Composite & 1673 & 4.1\\
Passive & 744 & 1.8\\
\hline
\end{tabular}%
}
\label{tab:1} 
\end{table}
\vspace{-0.3cm}

\begin{figure}
\begin{center}
\hbox{
\includegraphics[clip, trim=0 0.01cm 0 0.2cm, height=5cm]{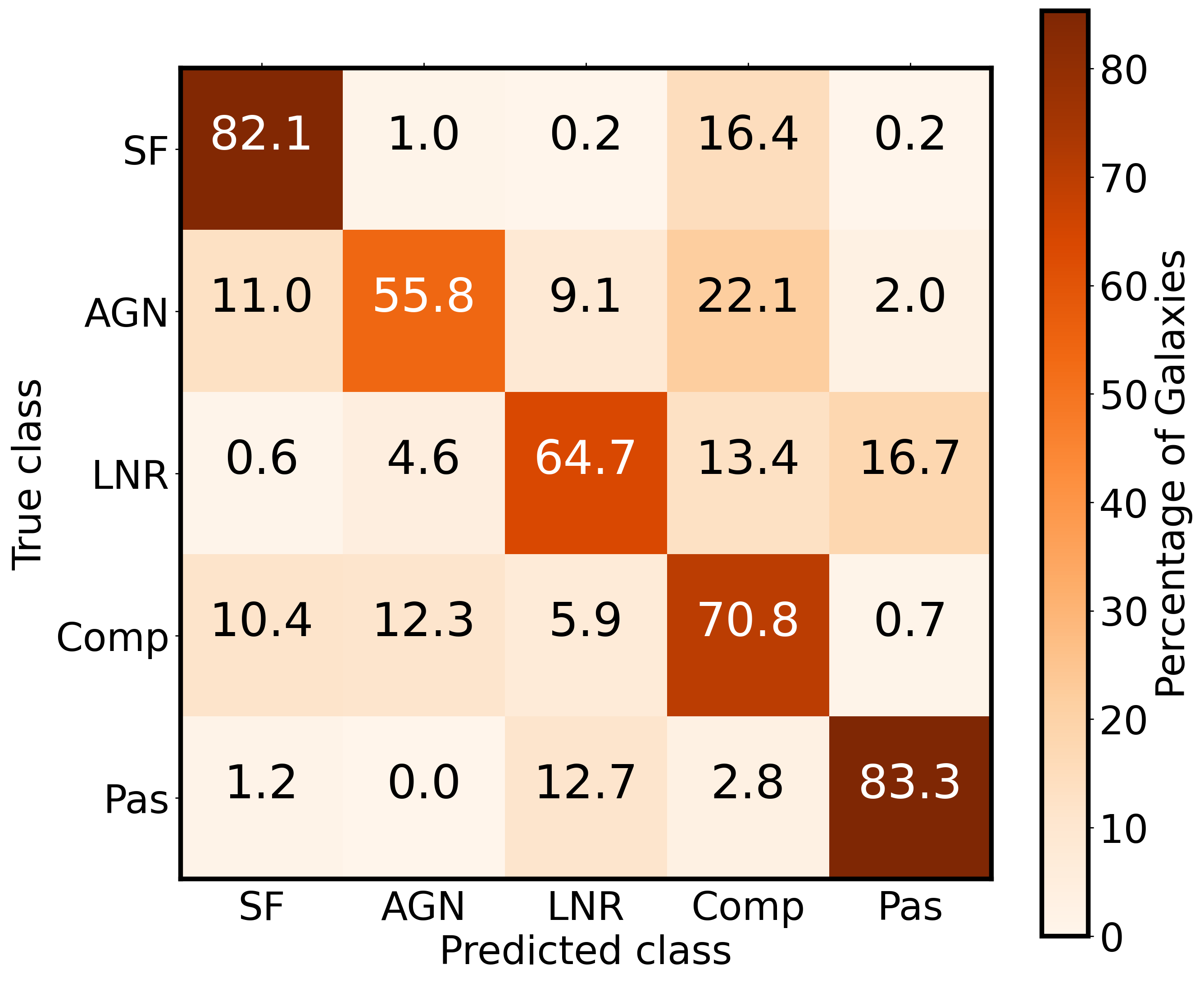}
}
\end{center}
\vspace{-0.75cm}
\caption{ \footnotesize 
 The confusion matrix of our classifier. The completeness for the SF and passive galaxies is very high ( $>$80\%) while for the other 3 classes it is lower as expected given the strong mixing between them in the feature distribution.
}
\label{fig:2}
\end{figure}

\begin{figure}
\begin{center}
\hbox{
\includegraphics[clip, trim=0 0.1cm 0 0.2cm, height=5cm]{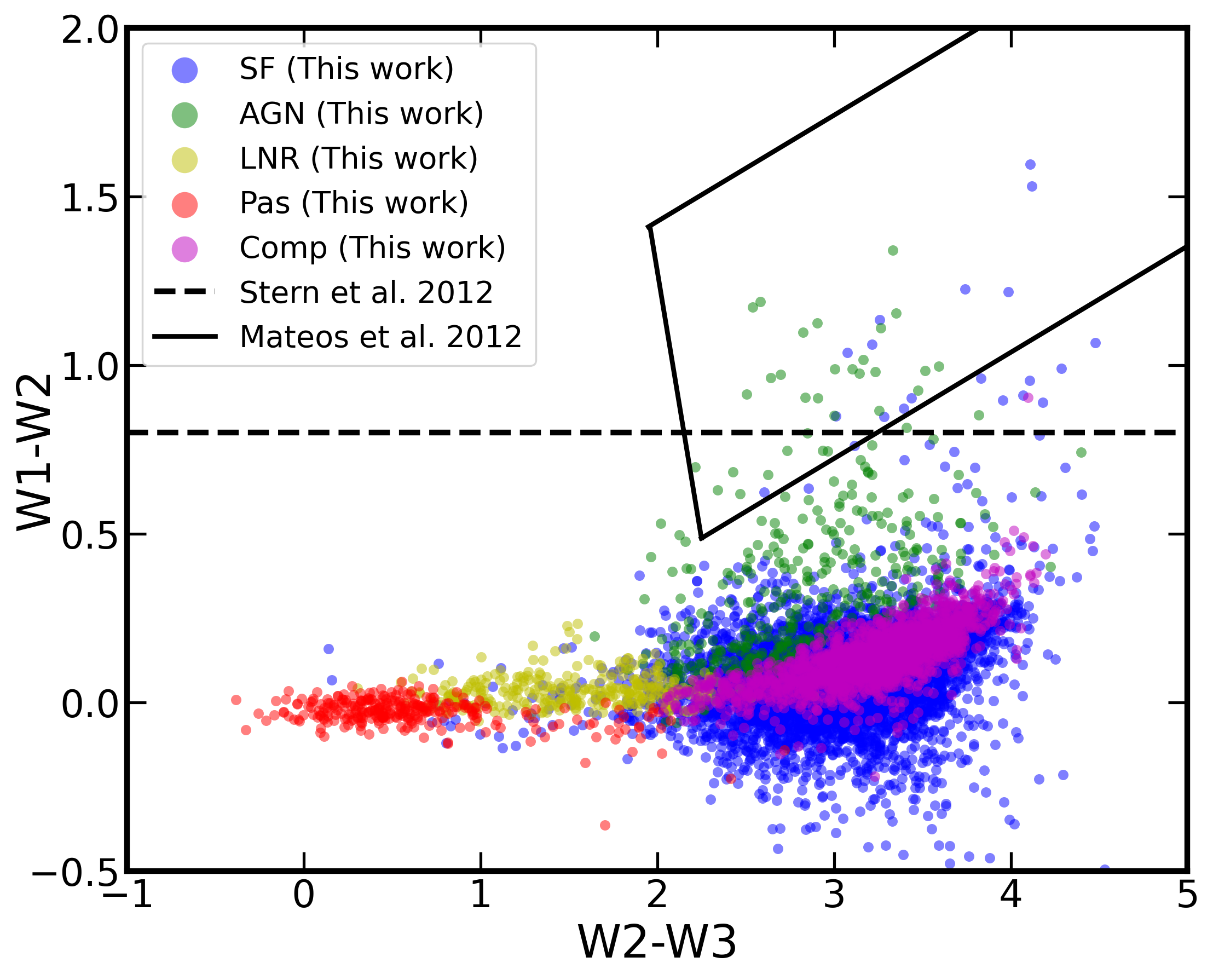}
}
\end{center}
\vspace{-0.75cm}
\caption{ \footnotesize 
 Comparison between our classifier and the standard mid-IR diagnostics applied on the HECATE catalog. Our method shows a population of AGN outside the AGN locus in the other mid-IR diagnostics, increasing the completeness towards lower luminosity AGN.
}
\label{fig:3}
\end{figure}

\section{Results}
The confusion matrix (Fig. \ref{fig:2}) shows that our classifier reaches maximum completeness of $\sim$82\% for SF and Passive galaxies and $\sim$56\% for AGN. 
This performance is expected if we consider the feature distribution of our training sample where the 5 classes are reasonably separated with higher mixing between the composite and AGN galaxies (Fig. \ref{fig:1}). Furthermore, these high scores indicate the robustness and reliability of our classifier when it is applied to unseen data (i.e. test set). 

We apply our new diagnostic to the HECATE nearby galaxy catalog (D$\leq$200 Mpc) \citep{Kovlakas2021}, and we compare our classifications with the mid-IR diagnostics from M13 and A12. 
Our new classifier reveals a large population of AGN outside their locus as defined in the other mid-IR diagnostics (green points below the dashed line in Fig.\ref{fig:3}). In particular, in a sample of 1227 spectroscopically classified AGN we find that our method recovers $\sim36 \%$ of the initial sample, while the M13, and A12 methods recover $\sim5 \%$ and $\sim6 \%$, respectively. Thus our new diagnostic increases the completeness of AGN identified with mid-IR colors since the other methods are more sensitive to luminous AGN, omitting a significant fraction of lower luminosity AGN. The reason for the success of our method is that the inclusion of the optical color allows the classifier to identify more AGN and also cases of starbursts with extreme mid-IR colors that mimic obscured AGN galaxies (blue points at the top right of Fig. \ref{fig:3}).

\vspace{-0.3cm}
\bibliographystyle{aa}
\bibliography{bonifacio}

\end{document}